\documentclass[11pt]{article}

\usepackage{graphicx}
\usepackage{epstopdf}
\usepackage{amsmath}
\usepackage{amssymb}
\usepackage{amsfonts}
\usepackage{amsthm}
\usepackage[usenames]{color}
\usepackage{array}
\usepackage{physics}

\usepackage[
colorlinks=true,
linkcolor=blue,
urlcolor=blue,
filecolor=blue,
citecolor=red,
pdfstartview=FitV,
pdftitle={},
pdfauthor={},
pdfsubject={},
pdfkeywords={},
pdfpagemode=None,
bookmarksopen=true
]{hyperref}

\usepackage{epsfig}

\usepackage{hyperref}

\newcommand{\bea}{\begin{eqnarray}}
\newcommand{\eea}{\end{eqnarray}}
\newcommand{\ba}{\begin{array}}
	\newcommand{\ea}{\end{array}}
\newcommand{\ee}{\end{equation}}

\numberwithin{equation}{section}

\begin{document}

\begin{flushright}
	\texttt{\today}
\end{flushright}

\begin{centering}
	
	\vspace{2cm}
	
	\textbf{\Large{
			  Flat-Space Limit of Holographic Pseudo Entropy \\ in (A)dS Spacetimes  }}
	
	\vspace{0.8cm}
	
	{\large   Reza Fareghbal }
	
	\vspace{0.5cm}
	
	\begin{minipage}{.9\textwidth}\small
		\begin{center}
			
			{\it  Department of Physics,
				Shahid Beheshti University, 1983969411,
				 Tehran , Iran \\}
				  			
			\vspace{0.5cm}
			{\tt   r$\_$fareghbal@sbu.ac.ir, reza.fareghbal@gmail.com}
			\\
			
					\end{center}
	\end{minipage}


	\begin{abstract}
	The real part of pseudo entropy in conformal field theories is  holographically calculated   by the area of some extremal spacelike surfaces in the  dual  dS and AdS spacetimes. We show that the flat-space limit of these curves in three-dimensional  (A)dS spacetimes  is well defined. We find that if the length of the curves  is calculated from the radial coordinate where the retarded time is extremum, then after taking the flat-space limit, the entanglement entropy of the dual theory of three-dimensional flat spacetime is obtained. For dS spacetime, the radial coordinate corresponding to the extremum  of retarded time is located inside the cosmological horizon. Our results suggest  that on the  field theory side, the entanglement entropy in the  dual theory of flat spacetimes should be obtained from the ultrarelativistic limit of pseudo entropy in the dual CFT to (A)dS spacetimes.

	\end{abstract}

\end{centering}

\newpage



\section{Introduction}
 It has been proposed in \cite{Bagchi:2010zz}-\cite{Bagchi:2012cy} that the flat-space limit (i.e., the zero cosmological constant limit) of asymptotically AdS spacetimes is equivalent to the ultrarelativistic limit \cite{leblond} in conformal field theory. As a result, it has been suggested that the dual field theory to asymptotically flat spacetimes in three dimensions is a two-dimensional ultra-relativistic field theory, known as Carrollian Conformal Field Theory (CCFT) \cite{Bagchi:2012cy,Duval:2014uoa,Duval:2014uva}. This theory exists at null infinity of asymptotically flat spacetimes, and this duality is referred to as the flat/CCFT correspondence. \footnote{In some papers, this duality is also called flat/BMSFT. BMSFT refers to a field theory with BMS symmetry, which is the asymptotic symmetry of asymptotically flat spacetimes at null infinity \cite{Bondi:1962px}-\cite{Barnich:2009se}. This symmetry is equivalent to Carrollian conformal symmetry in one lower dimension.}

One notable work in the field of  AdS/CFT is the Ryu-Takayanagi proposal. According to this proposal, the entanglement entropy in conformal field theories (CFTs) corresponds to the area of extremal surfaces in asymptotically AdS spacetimes. These extremal surfaces are connected to the subsystem on the AdS boundary for which the entanglement entropy is calculated. This correspondence is significant because it simplifies the calculation of entanglement entropy in field theory by relating it to more straightforward calculations in gravity.

Typically, a general system is decomposed into subsystems at constant time levels. This allows the Hilbert space to be expressed as a tensor product of its subspaces. From a general density matrix in the original system, one can derive a reduced density matrix for spacelike subsystems and calculate the entanglement entropy. For CFTs, the holographic dual of this entanglement entropy, according to Ryu-Takayanagi \cite{Ryu:2006bv}, is represented by extremal spacelike surfaces in the bulk of asymptotically AdS spacetimes. Recently, the authors of \cite{Doi:2022iyj}-\cite{Doi:2023zaf} have studied the generalization of entanglement entropy for timelike subsystems and proposed holographic duals for these as well. This entropy, known as timelike entanglement entropy, is a complex quantity where the real part corresponds to the area of spacelike extremal surfaces and the imaginary part corresponds to the area of timelike extremal surfaces in the bulk. In asymptotically AdS spacetimes, these spacelike surfaces are divided into several parts connected to the subsystem at the boundary but not directly to each other in the bulk, with timelike surfaces connecting them. Furthermore, \cite{Doi:2022iyj}-\cite{Doi:2023zaf} argue that timelike entanglement entropy is essentially a special case of pseudo-entropy \cite{Nakata:2020luh}, where instead of the reduced density matrix, the transition matrix is used (see section \ref{psuedo t}).

The Ryu-Takayanagi proposal for holographic entanglement entropy can be explored in other dualities. For asymptotically de Sitter (dS) spacetimes, this correspondence is known as dS/CFT, suggesting that a (d-1)-dimensional Euclidean conformal field theory is the holographic dual of d-dimensional asymptotically dS spacetimes \cite{Strominger:2001pn}. This field theory, defined at the timelike infinity of dS spacetime, is non-unitary, so the entanglement entropy for its subsystems is essentially pseudo-entropy and must be complex. The holographic dual of this pseudo-entropy is given in \cite{Doi:2022iyj}-\cite{Doi:2023zaf}, consisting of extremal surfaces in the bulk, with parts being spacelike and others timelike (see also \cite{Narayan:2015vda}-\cite{Narayan:2023zen}). Unlike in AdS, the timelike surfaces here connect to the subsystem at timelike infinity, while the spacelike surfaces appear inside the bulk and connect to the timelike surfaces.

For CCFTs, as the dual theory of asymptotically flat spacetimes, entanglement entropy can also be defined. Using the infinite-dimensional symmetry of two-dimensional CCFTs, a universal formula for the entanglement entropy of intervals is provided in \cite{Bagchi:2014iea} (see also the recent work \cite{Banerjee:2024ldl}). The holographic interpretation of this entropy is proposed in \cite{Jiang:2017ecm} as extremal curves in asymptotically flat spacetimes. These curves consist of spacelike and null parts, with the spacelike part entirely in the bulk and the null part connecting the ends of the interval at null infinity to the spacelike curve.

The method used in \cite{Jiang:2017ecm} to find extremal surfaces generalizes Rindler's method \cite{Casini:2011kv} to flat/CCFT. By applying a unitary transformation known as the Rindler transformation, entanglement entropy is mapped to thermal entropy, whose holographic counterpart is the entropy of the cosmological horizon in three-dimensional asymptotically flat spacetimes, known as flat-space cosmology (FSC) \cite{Cornalba:2002fi}-\cite{Cornalba:2003kd} (see also \cite{Bagchi:2012xr}). This spacetime is obtained from the flat-space limit of the BTZ black hole and has only one cosmological horizon. The Rindler transformation is extended to the bulk, and its inverse relates the FSC horizon to the extremal surface, whose area is proportional to the entanglement entropy of subsystems in CCFT.

Given that asymptotically flat spacetimes are derived from the flat-space limit of asymptotically (A)dS spacetimes, it should be possible to connect extremal surfaces in these spacetimes with those in asymptotically flat spacetimes. It is known that the flat-space limit of the extremal surfaces introduced by Ryu and Takayanagi for AdS spacetime is not well-defined. This issue is addressed in \cite{Fareghbal:2020dtq}, which introduces new curves in three-dimensional AdS spacetime using the Rindler method. These curves map to the inner horizon of the BTZ black hole rather than the outer horizon. The flat limit of these curves leads to surfaces proposed in \cite{Jiang:2017ecm} as the holographic dual of entanglement entropy in CCFT.

In this paper, we demonstrate that the new curves in \cite{Fareghbal:2020dtq} are indeed the same as the spacelike part of the timelike entanglement entropy in the dual CFT to AdS. In other words, to obtain extremal curves in asymptotically flat spacetimes, one must use extremal curves corresponding to pseudo-entropy in CFT. We show that this interpretation holds for both AdS and dS spacetimes, and the spacelike part of the extremal curves, whose length corresponds to the real part of pseudo-entropy in the dual CFT, has a well-defined flat-space limit. Although the total length of the spacelike curves after the flat-space limit differs from the entanglement entropy of CCFT, we find that if we consider only the portion of these curves starting from the extremum of retarded time and extending to the end in the bulk, the length divided by 4G, where G is Newton's constant, exactly gives the entanglement entropy of CCFT.

In the case of dS spacetime, the spacelike curve before the flat-space limit begins from a point outside the cosmological horizon and stretches to the minimum radius inside the horizon after crossing it. Since the extremum of retarded time is located inside the horizon, a segment of the spacelike curve that corresponds to the entanglement entropy of CCFT lies entirely within the cosmological horizon.

Our calculations in this paper are entirely within the gravitational sector of duality. The fact that the flat-space limit of curves corresponding to pseudo-entropy in dS and AdS results in curves corresponding to entanglement entropy in flat spacetimes suggests a similar connection on the field theory side. In other words, to find entanglement entropy in CCFT by taking the ultra-relativistic limit from a quantity in CFT, this quantity should be pseudo-entropy rather than entanglement entropy.

\section{Preliminaries}

\subsection{Entanglement entropy and pseudo entropy in  field theory}\label{psuedo t}

In a quantum field theory, suppose  the Hilbert space ${\mathcal{H}}$ is factorized as ${\mathcal{H}}={\mathcal{H}}_A\otimes {\mathcal{H}}_B$. This means that  the total system is divided into two subsystems A and B, where B is the complement of A. These subsystems are assumed to be spacelike regions of the main system. If $\rho$ is the density matrix of the total system, then the reduced density matrix for subsystem A,  denoted by $\rho_A$, is obtained by tracing out B as 
\begin{equation}\label{rhoa}
\rho_A=\text{Tr}_B \rho.
\end{equation}
The entanglement entropy for  subsystem A, $S_A$, is  the von Neumann entropy  calculated using the reduced density matrix $\rho_A$, i.e.
\begin{equation}\label{SA}
S_A=-\text{Tr}\left[\rho_A\log \rho_A\right].
\end{equation}
If $\ket{\psi}$ and $\ket{\varphi}$ are two pure states of the total system, we can define a reduced transition matrix as follows:
\begin{equation}\label{transmat}
\tau_A=\text{Tr}_B\left[\dfrac{\ket{\psi}\!\bra{\varphi}}{\bra{\varphi}\ket{\psi}}\right].
\end{equation}
Using this transition matrix, pseudo entropy $S_A^{(p)}$ is defined as \cite{Nakata:2020luh}
\begin{equation}\label{Sp}
S_A^{(p)}=-\text{Tr}\left[\tau_A\log \tau_A\right].
\end{equation}

If subsystem A is a timelike region  of the total system,  the reduced density matrix can still be defined for it using formula \eqref{rhoa}, and the  entanglement entropy can be  obtained using formula \eqref{SA}. However, in this case, the entropy  is complex-valued. This entropy is referred to as timelike  entanglement entropy. The complexity arises because  the density matrix in this case is not Hermitian. In  the paper \cite{Doi:2023zaf}, it is argued that the timelike entanglement entropy can be reformulated  using a transition matrix. Therefore, the timelike entanglement entropy itself is considered  a  pseudo entropy.

If the field theory is non-unitary, then generally $\bra{\psi}\neq \ket{\psi}^\dagger$. Consequently,  the operator $\rho=\ket{\psi}\!\bra{\psi}$ behaves  more like a transition matrix than a  density  matrix. Thus, the von Neumann entropy  defined using this $\rho$ and its associated  reduced matrices, will  indeed be pseudo entropy.

\subsection{Entanglement entropy in the dual theory of flat spacetimes}

In  papers \cite{Bagchi:2010zz}-\cite{Bagchi:2012cy} , it is proposed that the dual field theory to  asymptotically flat spacetime in three dimensions is a two-dimensional CCFT, which possesses the following infinite-dimensional symmetry:
\begin{align}\label{algebra}
\nonumber & [L_n,L_m]=(n-m)L_{n+m}+\dfrac{c_L}{12}(n^3-n)\delta_{n+m,0},\\
\nonumber & [L_n,M_m]=(n-m)M_{n+m}+\dfrac{c_M}{12}(n^3-n)\delta_{n+m,0},\\
&[M_n,M_m]=0,
\end{align}
where $c_L$ and $c_M$ are central charges.

This algebra is derived from the  Inonu-Wigner contraction of the  conformal algebra in two dimensions. This contraction represents   the ultra-relativistic  limit \cite{leblond} of  CFT, which is why  the term Carrollian is used to describe   this field theory \cite{Duval:2014uoa,Duval:2014uva}. The main proposal  of \cite{Bagchi:2012cy} is that the  flat-space limit on the gravity side corresponds to the ultra-relativistic  limit in the dual CFT. 

It is evident from   algebra \eqref{algebra} that the symmetries of two-dimensional  CCFTs are infinite dimensional,  similar to those of  two-dimensional  CFTs . This infinite-dimensional symmetry allows for the identification of  universal properties within these  theories. Among these features are  universal formulas for entanglement entropy. Research into this aspect of  two-dimensional CCFTs began  in \cite{Bagchi:2014iea}, where it  was shown that,  in a CCFT on the plane with coordinates $(u,\phi)$ where $u$ is known as retarded time \cite{Barnich:2006av} and $\phi$ is non-periodic, if we consider the sub-system to be a line with end points $\left(u=-\dfrac{l_u}{2},\phi=-\dfrac{l_\phi}{2}\right)$ and  $\left(u=\dfrac{l_u}{2},\phi=\dfrac{l_\phi}{2}\right)$ where $l_u$ and $l_\phi$ are constants, the entanglement entropy is given by
\begin{equation}\label{ee plane}
S_{EE}=\dfrac{c_L}{6}\log\dfrac{l_\phi}{\epsilon}+\dfrac{c_M}{6}\dfrac{l_u}{l_\phi},
\end{equation}
 where  $\epsilon$ is a cutoff used to regulate the interval and $c_L$ and $c_M$ are central charges of \eqref{algebra}.
For  CCFTs on the cylinder with coordinates $(u,\phi)$ where $\phi$ is periodic, this formula is given by
\begin{equation}\label{ee cylinder}
S_{EE}=\dfrac{c_L}{6}\log\left(\dfrac{2}{\epsilon}\sin\dfrac{l_\phi}{2}\right)+\dfrac{c_M}{12}l_u\cot\dfrac{l_\phi}{2}.
\end{equation}

The holographic interpretation of entanglement entropy was proposed in \cite{Jiang:2017ecm}. The  algebra \eqref{algebra} on the gravity side corresponds to  the  asymptotic symmetry of asymptotically flat spacetimes  at  null infinity \cite{Barnich:2006av}. Consequently, dual CCFTs must be defined  at null infinity, and each subsystem in these theories   will correspond to intervals at null infinity as well. This implies that the   extremal spacelike curves , whose length leads to the entropies \eqref{ee plane} and \eqref{ee cylinder} , do not connect  the two ends of the interval directly. In fact, the proposal of \cite{Jiang:2017ecm} is that the holographic interpretation in this context involves spacelike   curves whose end points  are connected to the two ends of the interval  at  null infinity  by two null rays.

\section{Holographic pseudo entropy in   AdS  and its flat limit}
In this section we aim to find  the holographic timelike entanglement entropy for an  interval on the boundary of  AdS$_3$ and then take the flat-space limit. As an example of pseudo entropy in CFT, this entropy is inherently complex-valued . The holographic dual of this entropy  introduced in \cite{Doi:2022iyj}-\cite{Doi:2023zaf}  consists of two  geodesics: one spacelike and one timelike. To obtain the flat-space limit, we only need the spacelike part.  We  begin by considering a spacelike  interval on the boundary of AdS$_3$, compute its holographic entanglement entropy, and then perform a Wick rotation   to extract  the spacelike part of the holographic timelike entanglement entropy.
\subsection{Holographic pseudo entropy from holographic entanglement entropy}
Let us consider $AdS_3$ written in the Poincare coordinate,
\begin{equation}\label{Poincare}
ds^2=\dfrac{\ell^2}{z^2}\left(-dt^2+dx^2+dz^2\right),
\end{equation}
where $\ell$ is the AdS radius and the boundary is located at $z=0$. First, we  compute  the holographic entanglement entropy for a spacelike  interval in the dual CFT$_2$ residing on the boundary of this geometry. This interval is characterized by
\begin{equation}
-\dfrac{l_x}{2}<x<\dfrac{l_x}{2},\qquad -\dfrac{l_t}{2}<t<\dfrac{l_t}{2}
\end{equation}
where $l_x$ and $l_t$ are two constants with $l_x>l_t$, ensuring that  this interval is spacelike.  The entanglement entropy of this interval as a subsystem of the boundary CFT,  is given by a universal formula  that  depends only  on the length  of it and the  central charges of the CFT \cite{Calabrese:2004eu}. According to the  Ryu-Takayanagi proposal \cite{Ryu:2006bv}, this entanglement entropy has a holographic description as the length of an extremal curve in the bulk spacetime,  anchored to the two ends of the interval. In our case, this extremal curve is a semicircle given by
\begin{align}\label{RT in poincare}
\nonumber x^2=\dfrac{l_x^2\left(l_x^2-l_t^2-4z^2\right)}{4\left(l_x^2-l_t^2\right)},\\
t^2=\dfrac{l_t^2\left(l_x^2-l_t^2-4z^2\right)}{4\left(l_x^2-l_t^2\right)}.
\end{align}

Our main goal is to take the flat-space limit,  $\ell\to\infty$,  from the calculations and compare the results with those known  for asymptotically flat spacetimes. However, the  Poincare coordinate \eqref{Poincare} is not  suitable  for this limit. Therefore, we write the AdS$_3$ metric in  global BMS coordinate \cite{Barnich:2012aw} as 
\begin{equation}\label{BMS-global}
  ds^2=-\left(1+\dfrac{r^2}{\ell^2}\right)du^2-2dudr+r^2d\phi^2,
  \end{equation}  
which is given by the following transformation from the Poincare coordinate \eqref{Poincare}:
\begin{align}\label{coordinate change}
\nonumber z&=\dfrac{2\ell^2}{\Delta},\\
\nonumber t&=\dfrac{2\ell\left(r\sin{\dfrac{u}{\ell}}-\ell\cos{}\dfrac{u}{\ell}\right)}{\Delta},\\
\nonumber x&=\dfrac{2\ell r \sin{\phi}}{\Delta},\\
 \Delta&=\ell\sin{\dfrac{u}{\ell}}+r\cos{\dfrac{u}{\ell}}+r\cos{\phi}.
 \end{align} 
In this coordinate system, $u$ is known as the retarded time .
If we denote the boundary interval by $-\dfrac{l_u}{2}<u<\dfrac{l_u}{2}$ and  $-\dfrac{l_\phi}{2}<\phi<\dfrac{l_\phi}{2}$ then we find that 
\begin{align}\label{lx to lu}
\nonumber l_x&=\dfrac{4\ell\sin{\dfrac{\l_\phi}{2}}}{\cos{\dfrac{l_\phi}{2}}+\cos{\dfrac{l_u}{2\ell}}},\\
l_t&=\dfrac{4\ell\sin{\dfrac{\l_u}{2\ell}}}{\cos{\dfrac{l_\phi}{2}}+\cos{\dfrac{l_u}{2\ell}}}.
\end{align}
Now we can use  \eqref{coordinate change} and \eqref{lx to lu} to write \eqref{RT in poincare} in the global BMS coordinate \eqref{BMS-global}. After some simplification, we obtain the following equations:
\begin{align}\label{RT in the bmsglobal}
\nonumber & r=\dfrac{\ell\cos\left(\dfrac{l_\phi}{2}\right)\sin\left(\frac{u}{\ell}\right)}{\cos\left(\dfrac{l_u}{2\ell}\right)\cos(\phi)-\cos\left(\dfrac{l_\phi}{2}\right)\cos(\dfrac{u}{\ell})},\\
& r^2\sin^2(\phi)\sin^2\left(\dfrac{l_u}{2\ell}\right)=\sin^2\left(\dfrac{l_\phi}{2}\right)\left(\ell\cos\left(u\over\ell\right)-r\sin\left(\dfrac{u}{\ell}\right)\right)^2.
\end{align}
It is clear that the flat-space limit ( $\ell\to\infty$ ) is not well-defined for the second equation \eqref{RT in the bmsglobal}. Therefore, the flat-space limit from the extremal curves given by using the Ryu-Takayanagi method for the spacelike interval  is not well-defined. This implies  that the holographic description of the entanglement entropy of  CCFT$_2$ (introduced in \cite{Jiang:2017ecm}) can not be obtained by taking the flat-space limit of \eqref{RT in the bmsglobal}. Our main goal is to find  appropriate extremal curves with a well-defined flat-space limit. The proposal is that these curves are the holographic dual of timelike entanglement entropy. Precisely  , the extremal curve with a well-defined flat space limit  is obtained by interchanging $\ell \l_\phi\leftrightarrow \l_u$ in \eqref{RT in the bmsglobal}. The equations describing our new extremal curve are 
\begin{align}\label{new extremal curve bmsglobal}
\nonumber & r=\dfrac{\ell\cos\left(\dfrac{l_u}{2\ell}\right)\sin\left(\frac{u}{\ell}\right)}{\cos\left(\dfrac{l_\phi}{2}\right)\cos(\phi)-\cos\left(\dfrac{l_u}{2\ell}\right)\cos(\dfrac{u}{\ell})},\\
& r^2\sin^2(\phi)\sin^2\left(\dfrac{l_\phi}{2}\right)=\sin^2\left(\dfrac{l_u}{2\ell}\right)\left(\ell\cos\left(u\over\ell\right)-r\sin\left(\dfrac{u}{\ell}\right)\right)^2.
\end{align}
This new curve is also  spacelike and  intersects the boundary  at two points $\left(u=\pm\frac{\ell l_\phi}{2},\phi=\pm\frac{l_u}{2\ell}\right)$. These  points mark the ends of a timelike interval on the boundary of AdS$_3$. This interval represents the timelike (Wick-rotated)   counterpart  of the original spacelike interval  considered as the subsystem in the dual CFT. Comparing \eqref{new extremal curve bmsglobal} with the results of \cite{Doi:2022iyj}-\cite{Doi:2023zaf} indicates  that our new curve is precisely  the spacelike part of the holographic timelike entropy.  The flat-space limit of this curve is
\begin{align}\label{flat extremal curve}
\nonumber &r=\dfrac{u}{\cos\phi\cos\left({l_\phi\over2}\right)-1},\\
&r^2\sin^2(\phi)\sin^2\left(\frac{l_\phi}{2}\right)=\dfrac{l_u^2}{4}.
\end{align}
which is consistent with the results of  \cite{Jiang:2017ecm}.

The equations of new curve \eqref{new extremal curve bmsglobal}, can be rewritten as 
\begin{align}\label{eo new cureve rewritten}
\nonumber &\cos \phi=\dfrac{\cos\left(\dfrac{l_u}{2\ell}\right)\left(r\cos\dfrac{u}{\ell}+\ell\sin\dfrac{u}{\ell}\right)}{r\cos\dfrac{l_\phi}{2}},\\
& \cos^2\phi=\cos^2\dfrac{l_u}{2\ell}+\dfrac{\ell^2\sin^2\dfrac{l_u}{2\ell}\cos^2\dfrac{l_u}{2\ell}}{r^2\left(\sin^2\dfrac{l_u}{2\ell}-\sin^2\dfrac{l_\phi}{2}\right)}.
\end{align}
Note that we still have the condition $\sin\dfrac{l_\phi}{2}>\sin{\dfrac{l_u}{2\ell}}$ which ensures that  the new curve remains spacelike. From the second equation of \eqref{eo new cureve rewritten}, we find that the radial coordinate $r$ has a minimum value which we denote as $r_{min}$ and 
\begin{equation}\label{def of r}
r_{min}=\dfrac{\ell\sin\dfrac{l_u}{2\ell}}{\sqrt{\sin^2\dfrac{l_\phi}{2}-\sin^2\dfrac{l_u}{2\ell}}}.
\end{equation}
Moreover, our new extremal curve consists of two disconnected branches.
Using the definition of $r_{min}$ \eqref{def of r}, we can write \eqref{eo new cureve rewritten} in its most simplified form,
\begin{align}\label{most simplified }
\nonumber &\cos \phi=\dfrac{\cos\left(\dfrac{l_u}{2\ell}\right)\left(r\cos\dfrac{u}{\ell}+\ell\sin\dfrac{u}{\ell}\right)}{r\cos\dfrac{l_\phi}{2}},\\
& \cos(\phi)=\pm\dfrac{\cos\dfrac{l_u}{2\ell}}{r}\sqrt{r^2-r_{min}^2},
\end{align}
where $\pm$ shows the equation of two branches. 
Combining two equations of \eqref{most simplified } we can omit $\phi$ and find
\begin{equation}\label{u in terms of r}
r\cos\frac{u}{\ell}+\ell\sin\dfrac{u}{\ell}=\pm\cos\dfrac{l_\phi}{2}\sqrt{r^2-r_{min}^2}.
\end{equation}
Using this equation we can determine the extremum  of $u$. For both  branches, these  values  occur  at the  same radial coordinate $r=r_I$ where
\begin{equation}\label{def of rstar}
r_I=\dfrac{\ell\sin\dfrac{l_u}{2\ell}\cos\dfrac{l_u}{2\ell}}{\sin^2\dfrac{l_\phi}{2}-\sin^2\dfrac{l_u}{2\ell}}.
\end{equation}

\subsection{Calculation of length and taking flat-space limit}
To find the lengths of the two branches, we first write them in the Poincare coordinate \eqref{Poincare}. From \eqref{lx to lu} we observe that exchanging  $l_u\leftrightarrow \ell l_\phi$ in the BMS-global coordinate is equivalent to exchanging $l_x\leftrightarrow l_t$ in the Poincare coordinate. Therefore, we can write our new extremal curve \eqref{new extremal curve bmsglobal} in terms of  the Poincare coordinate  by starting from \eqref{RT in poincare} and performing the exchange $l_x\leftrightarrow l_t$,
\begin{align}\label{new cureve  in poincare}
\nonumber x^2=\dfrac{l_t^2\left(l_t^2-l_x^2-4z^2\right)}{4\left(l_t^2-l_x^2\right)}\\
t^2=\dfrac{l_x^2\left(l_t^2-l_x^2-4z^2\right)}{4\left(l_t^2-l_x^2\right)}.
\end{align}
 The condition $l_x>l_t$ is again imposed to ensure that  this curve remains  spacelike for all values of $z$. We see that where  $r=r_{min}$ or equivalently $\cos\phi=0$ in the global BMS coordinates, it  results in $\Delta=0$ in \eqref{coordinate change} which corresponds to $\{z=\infty, t=\infty, x=\infty\}$. Furthermore, $r=r_I$ in the global BMS coordinate  maps to $z=z_I$, where
\begin{equation}\label{def of zstar}
 z_I=\dfrac{2\ell^2}{\left(\cos\dfrac{l_\phi}{2}+\cos\dfrac{l_u}{2}\right)\sqrt{r_I^2-r_{min}^2}}. 
 \end{equation} 
 Using \eqref{lx to lu}, \eqref{def of r} and \eqref{def of rstar}, we can write \eqref{def of zstar} as
 \begin{equation}\label{smplified z star}
  z_I=4\ell\sqrt{64\ell^2l_x^2+\left(16\ell^2-l_x^2+l_t^2\right)^2}\left(\dfrac{2\ell}{l_t\left(16\ell^2-l_x^2+l_t^2\right)}-\dfrac{1}{8\ell l_x}\right).
  \end{equation} 
  Using \eqref{Poincare} and \eqref{new cureve  in poincare} we can simplify the line element as 
  \begin{equation}\label{line element ads in poincare}
  ds=\dfrac{\ell A dz}{z\sqrt{z^2+A^2}},
  \end{equation}
  where
  \begin{equation}\label{Def of A}
  A=\dfrac{\sqrt{l_x^2-l_t^2}}{2}.
  \end{equation}
  We can use \eqref{coordinate change} to write \eqref{line element ads in poincare}  in terms of the parameters of the global BMS coordinate \eqref{BMS-global},
\begin{equation}\label{line in bms ads}
  ds=\dfrac{-\ell d\Delta}{\sqrt{\Delta^2+\dfrac{4\ell^4}{A^2}}}.
  \end{equation}  
  From \eqref{coordinate change}, \eqref{eo new cureve rewritten} and \eqref{def of r}, we see that   $\Delta$ is zero when $r=r_{min}$. Thus, integrating  \eqref{line in bms ads} from an arbitrary radius $r$ corresponding to $\Delta=\Delta_i$ to $r=r_{min}$ or $\Delta=0$, gives  the following length for the two branches:
  \begin{equation}\label{Length of two branches}
      L=2\ell\sinh^{-1}\left(\dfrac{A\Delta_i}{2\ell^2}\right).
      \end{equation}  
 If $\Delta_i$  corresponds to the boundary of spacetime, $\dfrac{L}{4G}$ is exactly the spacelike  entanglement entropy of the original spacelike interval or the real part of the timelike entanglement entropy of the timelike  interval. However, the flat-space limit of this quantity is not well-defined. It can be verified that a well-defined flat space limit is achievable  when $\Delta_i$ corresponds to $r=r_I$ where $r_I$ is given by \eqref{def of rstar}. In this case, by considering only  a portion of the curves from $r_{min}$ to $r_I$, $\Delta_i=\Delta_I$ where
 \begin{equation}\label{def of delta star}
      \Delta_I=\dfrac{\ell\sin\dfrac{l_u}{2\ell}\cos\dfrac{l_\phi}{2}}{\cos\dfrac{l_u}{2\ell}-\cos\dfrac{l_\phi}{2}}
      \end{equation}     
  and
  \begin{equation}\label{final lentgh before flat limit for ads}
      L=2\ell\sinh^{-1}\left(\dfrac{\sin\dfrac{l_u}{2\ell}\cos\dfrac{l_\phi}{2}}{\sqrt{\cos^2\dfrac{l_u}{2\ell}-\cos^2\dfrac{l_\phi}{2}}}\right)
      \end{equation}    
 The flat-space limit of \eqref{final lentgh before flat limit for ads} is well-defined and corresponds exactly to the entanglement entropy of an interval in the dual theory of flat spacetime \cite{Bagchi:2014iea} . It is evident that \eqref{final lentgh before flat limit for ads} does not represent  the entanglement entropy of an interval in the original CFT. An  interpretation  of this quantity in the context of the two dimensional CFT remains unresolved in this paper.  Note that $L$ in \eqref{final lentgh before flat limit for ads} is not invariant under any boost applied to the boundary interval. For instance  if $l_u=0$, then $L=0$, whereas  for the boosted intervals $L\neq 0$ .

\section{Holographic pseudo entropy in dS and its flat-space limit}
To find extremal curves with a well-defined flat-space limit in dS spacetimes, we use the fact that the metric of  dS$_3$  can be obtained   from the metric of  AdS$_3$ by replacing $\ell^2\to-\ell^2$. Thus, we start from  dS$_3$  written in the BMS coordinate as
\begin{equation}\label{ds in BMS}
ds^2=-\left(1-\dfrac{r^2}{\ell^2}\right)du^2-2du dr+r^2 d\phi^2,
\end{equation}
where $\ell$ is the dS radius and the cosmological horizon is located at $r=\ell$. 
Instead of calculating extremal curves directly for this spacetime, we can use the results of previous section and replace $\ell$ with $i\ell$. However the RT curve \eqref{RT in the bmsglobal}, after this substitution,  still does not yield a  well-defined flat-space limit. Therefore, we need to construct an extremal curve with a well-defined flat-space limit  by starting from our new curve \eqref{new extremal curve bmsglobal} and substituting  $\ell$ with $i\ell$. We propose the following extremal curve as the one with a well-defined flat space limit in dS spacetimes:
\begin{align}\label{new extremal curve ds }
\nonumber & r=\dfrac{\ell\cosh\left(\dfrac{l_u}{2\ell}\right)\sinh\left(\frac{u}{\ell}\right)}{\cos\left(\dfrac{l_\phi}{2}\right)\cos(\phi)-\cosh\left(\dfrac{l_u}{2\ell}\right)\cosh(\dfrac{u}{\ell})},\\
& r^2\sin^2(\phi)\sin^2\left(\dfrac{l_\phi}{2}\right)=\sinh^2\left(\dfrac{l_u}{2\ell}\right)\left(\ell\cosh\left(u\over\ell\right)+r\sinh\left(\dfrac{u}{\ell}\right)\right)^2.
\end{align}

To find the extension of this curve into the bulk, we first simplify  \eqref{new extremal curve ds } to
\begin{align}\label{simplified extremal ds}
\nonumber & \cos \phi=\dfrac{\cosh\dfrac{l_u}{2\ell}\left(\ell\sinh\dfrac{u}{\ell}+r\cosh\dfrac{u}{\ell}\right)}{r\cos\dfrac{l_\phi}{2}},\\
& \cos^2\phi=\cosh^2\dfrac{l_u}{2\ell}-\dfrac{\ell^2\sinh^2\dfrac{l_u}{2\ell}\cosh^2\dfrac{l_u}{2\ell}}{r^2\left(\sin^2\dfrac{l_\phi}{2}+\sinh^2\dfrac{l_u}{2\ell}\right)}.
\end{align}

From the second equation in \eqref{simplified extremal ds} we find that $r\geq r_{min}$ where
\begin{equation}\label{def of rmin ds}
r_{min}=\dfrac{\ell\sinh\frac{l_u}{2\ell}}{\sqrt{\sinh^2\dfrac{l_u}{2\ell}+\sin^2\dfrac{l_\phi}{2}}}.
\end{equation}
It is clear that $r_{min}<\ell$. Thus this extremal curve has a turning point at $r=r_{min}$ which is inside the cosmological horizon. Using   \eqref{def of rmin ds} we can write \eqref{simplified extremal ds} as
\begin{align}\label{ds extremal with rmin}
\nonumber & \cos\phi=\pm \cosh \dfrac{l_u}{2\ell}\sqrt{1-\dfrac{r_{min}^2}{r^2}},\\
& \ell\sinh\dfrac{u}{\ell}+r\cosh\dfrac{u}{\ell}=\pm r\cos\dfrac{l_\phi}{2}\sqrt{1-\dfrac{r_{min}^2}{r^2}}.
\end{align}
From the first equation in \eqref{ds extremal with rmin}, we conclude that our extremal curve is truncated at  a point where $r=r_{max}$ and
\begin{equation}\label{def of rmax}
r_{max}=\dfrac{\ell\cosh\dfrac{l_u}{2\ell}}{\sqrt{\sinh^2\dfrac{l_u}{2\ell}+\sin^2\dfrac{l_\phi}{2}}}.
\end{equation}
Note that $r_{max}>\ell$, and for our new extremal curve, $r_{min}<r<r_{max}$.  This curve obviously crosses the cosmological horizon. Moreover, the flat space limit of \eqref{ds extremal with rmin} matches the  results of \cite{Jiang:2017ecm}. Our new curve is precisely the spacelike part of the curve  proposed in \cite{Doi:2022iyj} as the holographic dual of pseudo entropy in dS/CFT. 

From the second equation of \eqref{ds extremal with rmin}, we can determine the radius at which $u$ is extremum. Denoting  this radius by $r_I$, and setting  $\dfrac{du}{dr}\Big|_{r=r_I}\!\!\!\!\!\!\!=0$, in the second equation of \eqref{ds extremal with rmin}, we obtain the following interesting equation:
\begin{equation}\label{def of r star ds}
r_I=\dfrac{r_{min}r_{max}}{\ell}.
\end{equation}
Using \eqref{def of rmin ds}, \eqref{def of rmax} and \eqref{def of r star ds}, we find that $r_I\leq \ell$. Thus a portion  of our extremal curve, characterized  by $r_{min}\leq r \leq r_{I}$, is always inside the cosmological horizon.

To find the line element in this case, it is sufficient to replace $\ell$ with $i\ell$ in \eqref{line in bms ads}. We obtain
\begin{equation}\label{line element ds}
ds=\dfrac{\ell d\Delta}{\sqrt{B^2-\Delta^2}},
\end{equation}
 where
 \begin{eqnarray}\label{delta B ds}
 \nonumber &\Delta=\ell\sinh\dfrac{u}{\ell}+r\cosh\dfrac{u}{\ell}+r\cos\phi,\\ & B^2=\dfrac{\ell^2\left(\cos\frac{l_\phi}{2}+\cosh\frac{l_u}{2\ell}\right)^2}{\sin^2\frac{l_\phi}{2}+\sinh^2\dfrac{l_u}{2\ell}}.
 \end{eqnarray}
Our curve will be spacelike if $B^2>\Delta^2$. Using the equations for the curve \eqref{ds extremal with rmin} along with  \eqref{delta B ds}, we find that this condition is satisfied for $r<r_{max}$. Therefore, our new extremal curve in dS spacetime is always spacelike. Moreover, using the equations for the curve \eqref{ds extremal with rmin} and the definition of $r_{min}$ and $r_{max}$ in \eqref{def of rmin ds} and \eqref{def of rmax}, we simplify \eqref{line element ds} as
\begin{equation}\label{line element ds final}
ds=\dfrac{\ell r dr}{\sqrt{\left(r^2-r_{min}^2\right)\left(r_{max}^2-r^2\right)}}.
\end{equation}
Since this extremal curve consists of two similar branches, the total length, $L$ of the curve is twice  the integral of \eqref{line element ds final} from $r_{min}$ to $r_{max}$. Evaluating   this integral yields, $L=\ell\pi$. Note that $L/4G$ is half of the entropy of the cosmological horizon and  is exactly  the real part of the pseudo  entropy calculated in \cite{Doi:2022iyj}.

While the  flat-space limit of \eqref{ds extremal with rmin}  is well-defined  ,the length diverges as $\ell\to\infty$. This indicates  that  to take the flat-space limit, one must  consider only  a portion of this curve. We find that this portion must be characterized by $r_{min}<r<r_I$,  where $r_I$ is given by \eqref{def of r star ds}. If we calculate the integral of \eqref{line element ds final} from $r_{min}$ to $r_I$, the length of the two branches will be
\begin{equation}\label{L star}
 L_I=2\ell\sin^{-1}\dfrac{\sinh\frac{l_u}{2\ell}\cos\frac{l_\phi}{2}}{\sqrt{\sinh^2\dfrac{l_u}{2\ell}+\sin^2\frac{l_\phi}{2}}}.
 \end{equation} 
A simple calculation gives
\begin{equation}
\dfrac{1}{4 G}\lim_{\ell\to\infty} L_I=\dfrac{l_u\cos\frac{l_\phi}{2}}{4G\sin \frac{l_\phi}{2}},
\end{equation}
which is precisely the entanglement entropy of CCFT \eqref{ee cylinder}  with $c_L=0$ and $c_M=\dfrac{3}{G}$.

 \section{Summary and future directions }

In this paper, we demonstrate that the extremal spacelike curves in 3D flat spacetime, whose lengths correspond to the entanglement entropy in CCFT$_2$, can be derived by taking the flat-space limit from the spacelike geodesics in dS and AdS spacetime. The length of these geodesics yields the real part of pseudo-entropy in CFT$_2$. Extending this relation to dual field theories suggests that the entanglement entropy in CCFT can be obtained from pseudo-entropy by taking the ultra-relativistic limit from CFT. Accepting this connection raises several questions that need to be addressed in future works. One such question is how  imaginary part of pseudo-entropy and its holographic dual, which are timelike geodesics, are canceled in the limiting process.

Another issue discussed in this article is that to accurately obtain the entanglement entropy in CCFT, one should calculate only the length of a segment of the spacelike curve rather than its entirety. This segment length is calculated from a point in dS or AdS spacetime where the time direction, $u$, is extremum. The reason for this and its significance in dual field theories are questions we aim to answer in future studies. This problem becomes more intriguing when we consider formulas \eqref{final lentgh before flat limit for ads} and \eqref{L star}, which result from calculating the segment length of spacelike curves before taking the flat-space limit. These two formulas, which can be transformed into each other by the $\ell\to i\ell$ transformation, are quantitative expressions in Lorentzian CFT and Euclidean CFT, respectively, whose flat-space limit yields the entanglement entropy of CCFT. Understanding this quantity and its relation to pseudo-entropy or entanglement entropy is also a compelling question that will be the focus of our future research.

\subsubsection*{Acknowledgment}
The author would like to thank M. Ghanbari for his contribution in the first stage of this work.


\end{document}